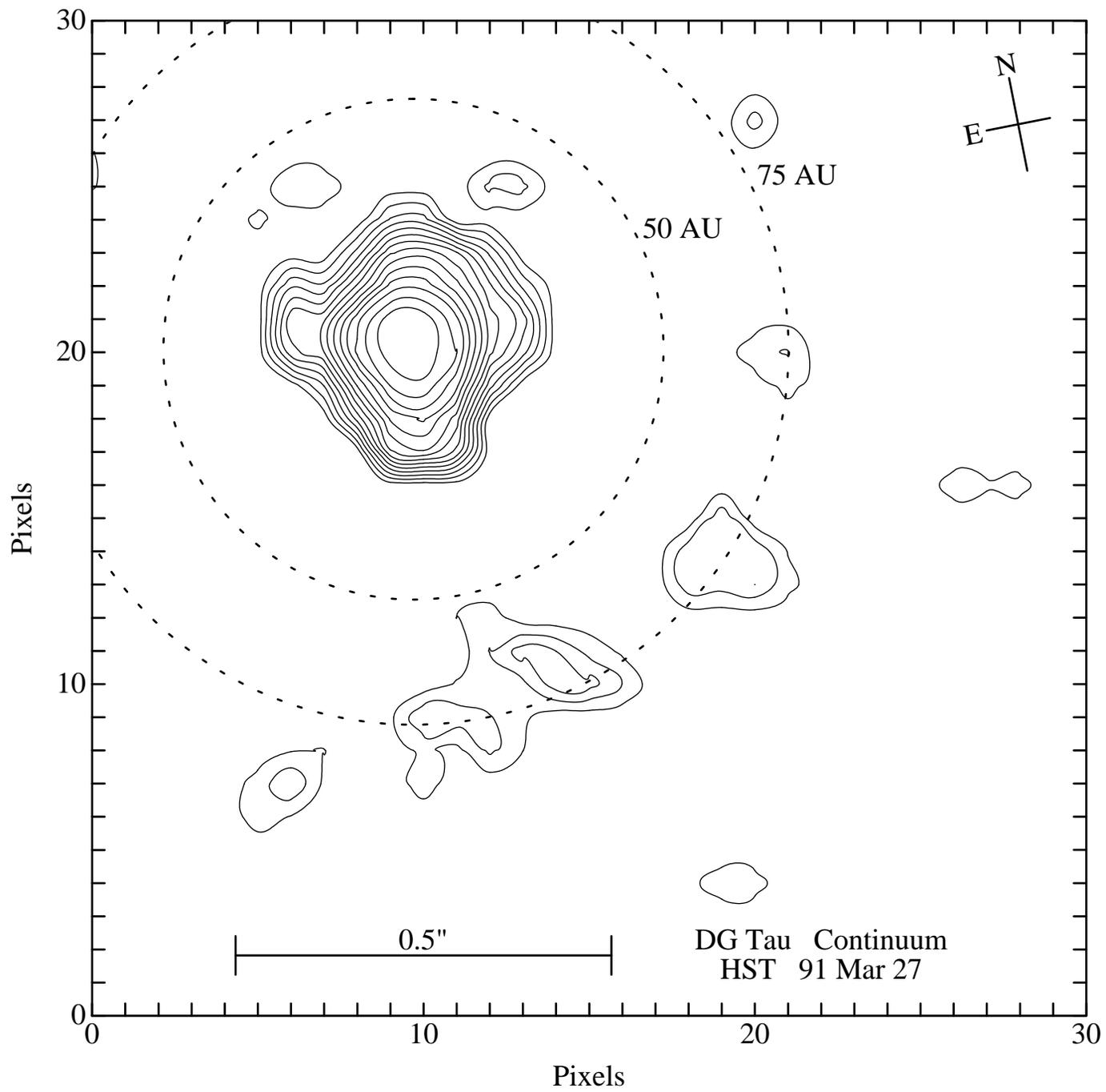

**Fig. 2c**

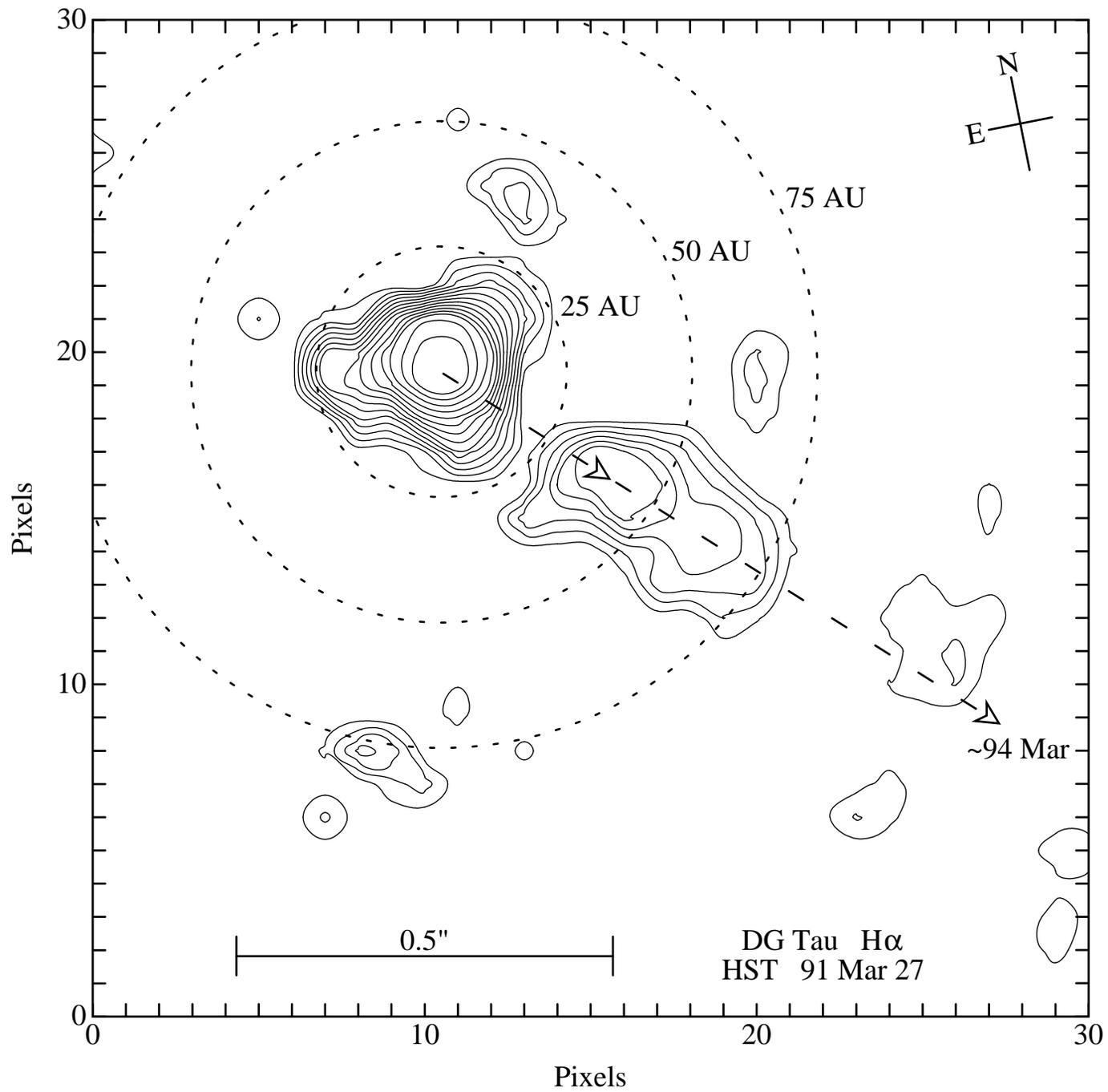

**Fig. 2a**

# Hubble Space Telescope Images of the Subarcsecond Jet in DG Tau[1]

J. Kepner and P. Hartigan

Five College Astronomy Department, University of Massachusetts, Amherst, MA 01003

C. Yang

Department of Statistics, University of Arizona, Tucson, AZ 85721

and

S. Strom

Five College Astronomy Department, University of Massachusetts, Amherst, MA 01003

## ABSTRACT

We have applied a new restoration technique to archival [O I], H$\alpha$, and continuum HST images of DG Tau. The restored [O I] and H$\alpha$ images show that DG Tau has a jet with a projected length of 25 AU and width $\leq$10 AU, and is already collimated at a projected distance of $\sim$ 40 AU (0$''$.25) from the star. Such a narrow width and short collimation distance for a stellar jet places important constraints on theoretical models of jet formation.

*Subject headings:* stars: individual (DG Tau) — stars: formation — stars: pre-main-sequence

## 1. Introduction

Bipolar molecular outflows and highly collimated optical jets emanate from many young stellar objects and affect the formation and evolution of young stars in several fundamental ways. Outflows limit the amount of material that accretes onto the protostar and its surrounding disk from the dark cloud; these flows also deposit large amounts of momentum and energy into the surrounding medium (Shu, Adams, & Lizano 1988; Margulis & Lada 1986). Stellar jets provide a way for the star to lose the angular momentum accreted from infalling disk material, allowing young stars to rotate slowly (Edwards, Mundt, & Ray 1993). Energy from an accretion disk

---

probably powers molecular outflows and stellar jets, but we do not know how young stars redirect infalling material from the disk into highly collimated supersonic bipolar outflows. It is even unclear as to whether or not a single physical mechanism produces both the molecular flow and the jet (Raga & Cabrit 1993; Kwan & Tademaru 1988). Current models of stellar jets and molecular outflows use the ambient medium or a magnetic field to collimate the flow (e.g. Canto, Tenorio-Tagle, & Rozyczka 1988; Konigl 1989), but existing observations constrain the models poorly because ground based images of stellar jets cannot determine if the jet is collimated closer to the star than several hundred AU. A recent study by Mundt, Ray, & Raga (1991) suggests that jets from young stars may become collimated at distances as large as 1000 AU from the star.

The bright T Tauri star DG Tau is one of the most active young stellar objects known, making it an excellent target for high resolution observations. Imaging with HST near its diffraction limited performance ($\sim 0\rlap{.}{''}1$ at optical wavelengths) offers the possibility of tracing stellar jets to within 10 AU of the stellar surface. DG Tau has a broad spectral energy distribution at infrared and millimeter wavelengths (Beckwith et al. 1990). Furthermore, the excess continuum at optical wavelengths exceeds the photospheric flux by a factor of $\sim 5$ (Hartigan et al. 1991). The above observations are consistent with an actively accreting disk. Rapid accretion through DG Tau's disk is probably responsible for driving an energetic mass outflow, which is seen as a faint jet extending $8''$ from the star (Mundt & Fried 1983). A 6-cm VLA image of DG Tau also shows an elongation in the direction of the optical jet, suggesting that the jet may be collimated on subarcsecond scales (Bieging, Cohen, & Schwartz 1984).

Although the aberration present in HST images has significantly reduced the ability of the telescope to study the collimation of stellar jets, the point spread function (PSF) of the HST is now sufficiently well-determined that it is possible to reconstruct reliable HST images. These recovered images have spatial resolution several times better than the best ground based observations. Archival HST images of DG Tau in the lines of [O I] and H$\alpha$ show extended emission along the direction of the jet that is not present in the point-spread function of the telescope. This hint of a resolved feature is the motivation for applying a new image restoration technique to clarify the morphology of the jet associated with DG Tau.

## 2. Observations

The observations of DG Tau, were made by the Wide Field/Planatery Camera (WF/PC) team on 1991 March 27. Four exposures of 400 seconds were taken with the Planetary Camera (HST ID: w0ia0602t, w0ia0603t, w0ia0604t, w0ia0605t), two each through the F656N H$\alpha$ ($\lambda = 6559$Å, $\Delta\lambda = 20$Å) and F631N [O I] ($\lambda = 6306$Å, $\Delta\lambda = 20$Å) narrow band filters. In addition, a single continuum exposure of 10 seconds (HST ID: w0ia0601t) was taken with the F702W ($\lambda = 6930$Å, $\Delta\lambda = 1470$Å) wide band filter. These configurations sampled the image at one pixel per $0\rlap{.}{''}044$. The data were flat fielded and calibrated at STScI using standard procedures. We obtained the images from the HST archive maintained by STScI in 1992 October.

Cosmic rays were removed using a gradient based algorithm. Unlike ground based observations, HST cosmic rays can have a very low angle of incidence, creating long lines of bad pixels. The gradient method handles these types of cosmic rays well, and can be applied to single images. The HST PSFs were calculated with the Tiny TIM (Telescope Image Modeling) program (Krist 1992; Krist 1991; Burrows & Hasan 1991). This program creates PSFs based on the camera, CCD position, filter, spectral type, and observation date. The advantage of Tiny TIM is that it generates a noiseless PSF quickly on most workstations. The disadvantage is that the PSF will have small differences with the "true" PSF, making it necessary to closely examine deconvolutions for artifacts.

The image restoration method used is based on the "Half Gaussian" model and the corresponding "Half Quadratic" algorithm. This newly developed method is described fully in Geman & Yang (1993), where comparisons are made with the widely used Lucy algorithm (Lucy 1974). A brief summary is given here. The observed data, $g$, are modeled as a random vector with mean $Hf$, where $f$ is the unknown idealized image, and $H$ is a matrix determined by the PSF. Because of noise and the ill-conditioned nature of $H$, the restoration method used here pursues an optimal compromise between the data and any prior knowledge about the object. Specifically, the restored image is defined as the global minimizer of the function:

$$\Phi(f) = \phi(Qf) + \lambda \|g - Hf\|^2$$

where $Q$ is a matrix and $\phi$ is a real-to-real function. The first term in the above expression imposes prior knowledge. The second term measures the distance between the data and the restored image, and the parameter $\lambda$ adjusts the balance between the two. The matrix $Q$ and nonlinear function $\phi$ allow prior knowledge to be included in the model. We used the "first-order smooth-discontinuity constraint" in our restorations (Geman & Reynolds 1992). This constraint measures the count differences of neighboring pixels and uses a concave $\phi$ to promote large smooth areas with a small number of sharp jumps.

Figure 1 shows the averaged original image (top), the PSF (middle), and the averaged restored imaged (bottom) of the H$\alpha$ and [O I] observations, respectively. The scale bar was computed assuming a distance of 150 pc which corresponds to a scale of $\sim$6.6 AU/pixel. Contour plots of the restored H$\alpha$, [O I], and continuum images are shown in Figure 2, adjacent contours represent a vertical change of a factor of two, and the circles indicate the radial distance from the star.

## 3. Discussion

An extension in the direction of the known DG Tau jet is visible in the original data, but the jet is masked by an arm of the PSF. Application of our image restoration technique reveals a knotty, jet-like feature emanating from the star along PA = 227° (see Figures 2a and 2b), as compared to PA = 226° from ground based observations in Mundt et al. (1987). The continuum image (Figure 2c) illustrates which features in the restored image are real and which are artifacts. For example, the circular arc at 75 AU is an artifact of the imperfect modeling of the PSF rings. These rings, which are highly wavelength dependent, also cause a faint arc at 60 AU in the [O I] image and make the star appear triangular in the H$\alpha$ and [O I] data. In general, the lowest three contours in Figure 2 are probably noise or artifacts of the deconvolution.

Figures 1 and 2 show that the jet is observed in both the raw and restored narrow band images and in neither of the continuum images. The deconvolved images distinctly show the separation between the star and the jet. While the jet is resolved along its length, there is no evidence that it is resolved along its width. The overall jet morphology shows little variation between the H$\alpha$ and [O I] images. The ratios of the total flux in the jet to the to the total flux from the star are nearly equal, 0.10 in H$\alpha$ versus 0.12 in [O I]. The parameters of the jet are identical in both the H$\alpha$ and [O I] frames: separation from the centroid of the star $\sim 0\farcs25$ (38 AU); length $\sim 0\farcs25$; width $\leq 0\farcs1$; length/width ratio $\geq 2.5$. The jet appears collimated, and because the width is unresolved, it is possible that the jet is even narrower than shown (r $\leq$ 10 AU). These data suggest that collimation takes place within a projected distance of 40 AU.

The most promising models for emission in jets are entrainment or time variability (e.g., Hartigan et al. 1993). Both mechanisms should produce motions of the knots in the jet that are visible with time. Ground-based observations of jets show knot proper motions on time scales of decades, but the high spatial resolution of HST allows measurement of knot proper motions on time scales of years. For example, if the measurements of Mundt et al. (1987) are valid close to the star and we take $v_{tan} \sim v_{rad} \sim$ 140 km/s (30 AU/yr), then the projected separation of the jet in 1994 April is estimated to be $\sim 1''$. Thus, the jet should show significant proper motion if it is re-imaged with HST. Additional observations of DG Tau with HST will determine the proper motion of the jet, constrain the projection angle of the flow and indicate the true collimation distance. Furthermore, the observation of new features will clarify the rate at which mass ejections occur.

### Note Added in Proof

New ground-based long slit spectra of DG Tau taken under excellent seeing conditions (Solf and Bohm, Ap J Letters in press) confirm that the jet emission located $\sim$ 0.3" from DG Tau has a large blueshifted radial velocity ($\sim$ -250 km/s).

The authors wish to thank Dr. Deborah Padgett for obtaining the HST images of DG Tau, Dr. Donald Geman for introducing us to his technique, Karen Strom for her helpful comments, and Dr. Karl Staplefeldt and David Wall for their help in analyzing the data. The work of J. Kepner was supported in part by NSF grant No. DMS-8813699, and by NASA through the HST GO program (No. 2265-01-87A) from the Space Telescope Science Institute, which is operated by the Association of Universities, Inc. under NASA contract NAS5-26555.

Fig. 1.— The left and right columns of images pertain to the H$\alpha$ and [O I] observations, respectively. Starting from the top, the images are: the averaged raw image, the HST PSF, and the averaged restored image. Note that the separation between the jet and the star, clearly visible in the restored image, is masked by an arm of the PSF in the raw image.

Fig. 2.— Contour plots of (a) H$\alpha$ (b) [O I] and (c) continuum restored images. Circles denote the distance from DG Tau in AU. The dashed line is the path of the outflow. The arrows denote the observed and predicted positions of the brightest portion of the jet.

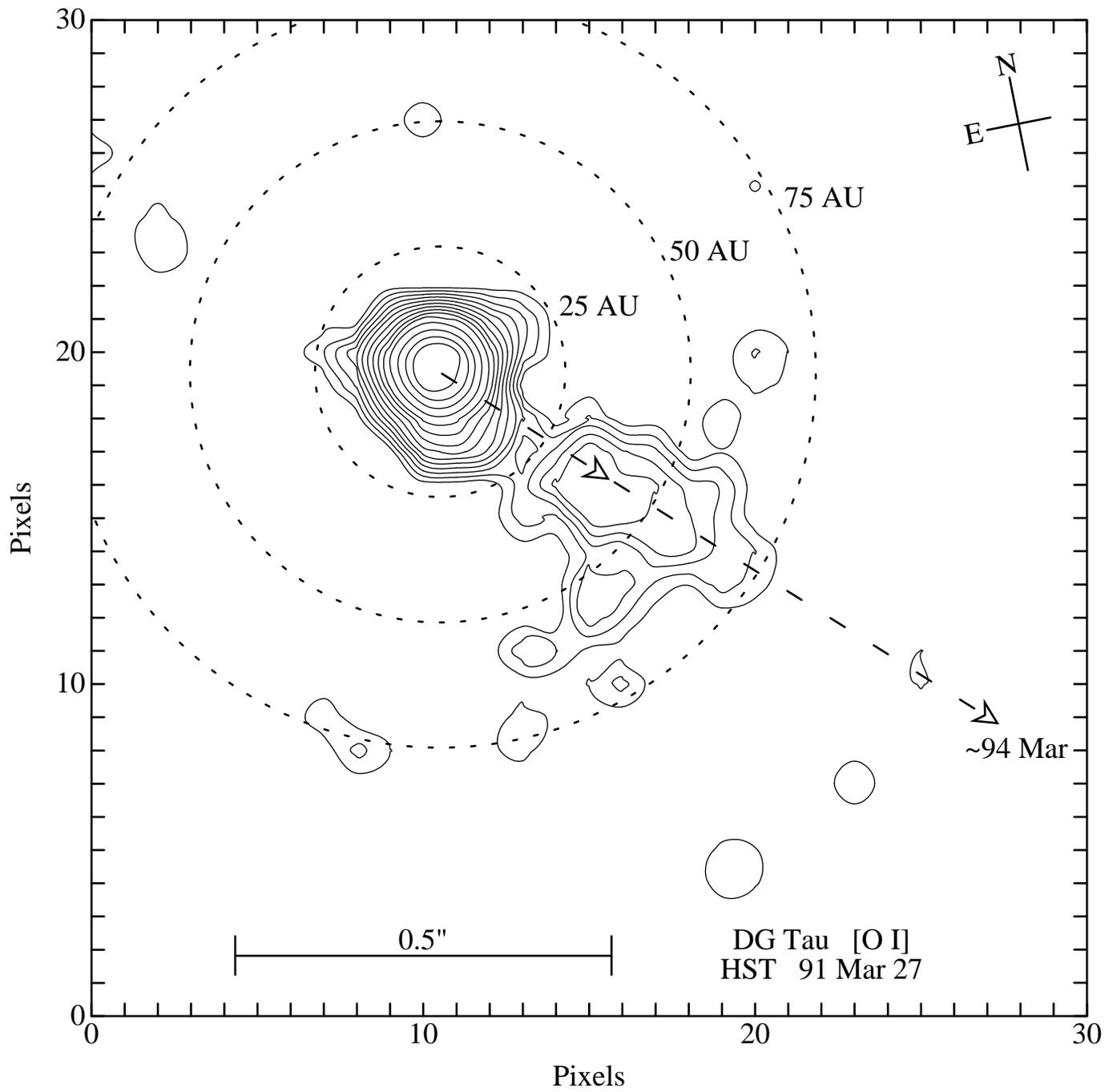

**Fig. 2b**